\documentclass[11pt,a4paper,twoside]{article}
\usepackage{amsmath}
\usepackage{amsfonts}
\usepackage{changes}
\numberwithin{equation}{section}

\begin{document}

\noindent

{\bf
{\Large Massive deformations of rank-2 symmetric tensor theory (a.k.a. BRS characterization of Fierz-Pauli massive gravity)
 
}} 

\vspace{.5cm}
\hrule

\vspace{1cm}

\noindent

{\large\bf{Alberto Blasi\footnote{\tt alberto.blasi@ge.infn.it }
and Nicola Maggiore\footnote{\tt nicola.maggiore@ge.infn.it }
\\[1cm]}}

\setcounter{footnote}{0}

\noindent
{{}Dipartimento di Fisica, Universit\`a di Genova,\\
via Dodecaneso 33, I-16146, Genova, Italy\\
and\\
{} I.N.F.N. - Sezione di Genova\\
\vspace{1cm}

\noindent
{\tt Abstract~:}
In this paper we consider the issue of massive gravity from a pure field theoretical point of view, as the massive deformation of the gauge theory for a symmetric rank-2 tensor field. We look for the most general massive theory with well defined propagators, imposing the absence of unphysical poles. We find several possibilities, depending on the choice of the gauge fixing term. {Amongst these, two solutions with good massless limit are found.} The request of the absence of \textcolor{black}{tachyons}, alone, does not isolate the Fierz-Pauli case: several examples of massive theories, which may include or not the Fierz-Pauli mass term,  are given. On the other hand, the Fierz-Pauli theory can be uniquely identified by means of a symmetry: it turns out the the Fierz-Pauli massive gravity is the only element of the cohomology of a BRS operator. 

\newpage

\section{Introduction}

In this paper we consider the theory defined on a $D$-dimensional flat Minkowski spacetime  for a symmetric rank-2 tensor $h_{\mu\nu}(x)$, and which is invariant under the gauge transformation 

\begin{equation}
\delta h_{\mu\nu}=\partial_\mu\theta_\nu + \partial_\nu\theta_\mu \label{1.1},
\end{equation}

where $\theta_\mu(x)$ is a local gauge parameter. In particular we are interested in the most general massive deformation which breaks the gauge symmetry \eqref{1.1}, possibly in a harmless way. 

What we have in mind, of course, is the modification of General Relativity (GR) at large distances, where $h_{\mu\nu}(x)$ is a perturbation of the metric around the flat Minkowski space $\eta_{\mu\nu}$, \textcolor{black}{in terms of which the linearized Einstein-Hilbert (EH) action is built}, but we do not necessarily restrict to that case. For a review on massive deformations of gravity see for instance  \cite{Rubakov:2008nh,Hinterbichler:2011tt,Comelli:2013txa,deRham:2014zqa}. Amongst the classical motivations for studying an infrared modified gravity, we would like to mention the recent interest in mechanisms breaking the diffeomorphism invariance of GR, coming from the holographic approach to the study of thermo-electric transport in condensed matter \cite{Blake:2013bqa,Amoretti:2014zha,Amoretti:2014mma}. There, the momentum dissipation results from the breaking of translational invariance due to impurities. Massive gravity represents a way to implement momentum relaxation in holography \cite{Vegh:2013sk}.

\textcolor{black}{
The boundaries within which we shall carry out our analisys are the following. We consider a symmetric, rank two tensor gauge field theory in $D$ spacetime dimensions whith flat Minkowski metric. According to the standard procedure in order to have well defined propagators we need to fix the gauge, which we shall do by means of a Lagrange multiplier vector field $b^{\mu}(x)$, and Faddeev-Popov ghost/antighost vector fields  $\xi^{\mu}(x)$ and $\bar\xi^{\mu}(x)$. The original gauge symmetry is implemented by the well known BRS invariance.}

\textcolor{black}{
In this approach the introduction of a mass term for the symmetric tensor field is viewed as a breaking of the gauge symmetry and of the BRS invariance as well. Hence the choice of a gauge is a necessary (not always sufficient) condition to have a good massless limit of the theory, \textcolor{black}{$i.e.$ well defined propagators for vanishing mass terms}.
}

\textcolor{black}{
We choose the most general gauge fixing term linear in the symmetric tensor field and the most general quadratic mass term, thus we end up with quite a few parameters: two for the mass sector and two for the gauge fixing. All of them are, so far, completely free.
}

\textcolor{black}{
In order to try to understand better the admissible range of these parameters we compute all the propagators of the theory, and require that there should be no ``unphysical poles'', $i.e.$ tachyons. There might be other sources of ``ghosts'', as the Boulware-Deser one \cite{Boulware:1973my}\footnote{\textcolor{black}{We remark that, in non linear massive gravity theory, the absence of the Boulware-Deser ghost has been proven in \cite{Hassan:2011hr}.}}. Anyway, since we only consider a quadratic theory, we stick with the minimal requirement of no tachyons. This is sufficient to limit the numerical range of the four parameters. Morevover there is a special case: in the Landau gauge and with the Fierz-Pauli (FP) choice of the mass term \cite{Fierz:1939ix} there is a modified version of the BRS symmetry which survives and uniquely identifies the corresponding action. Here the the situation is quite similar to massive QED, as pointed out in \cite{Baulieu:1983tg}. But, while in QED the symmetry tells nothing about the mass term, in our case it does fix the relative weight of the two mass parameters and it singles out the FP choice, which, being now linked to a symmetry, we expect to be protected.
}

%
  
In quantum field theory  methods exist, which allow to control mass breakings, for instance in supersymmetry, where ``soft'' supersymmetry breaking terms were first classified \cite{Girardello:1981wz}, and then identified as the only possible breaking terms  by means of a Slavnov-Taylor identity \cite{Maggiore:1996gg}. What we ask in this paper is whether something similar can be done in this case, $i.e.$ to characterize the FP massive gravity by means of a symmetry (BRS) tool.

Any deviation from the FP paradigm is mostly interesting, for instance for the realization of momentum dissipation  in gauge/gravity duality, and interesting efforts towards this direction have already been done, for instance by means of the introduction of Lorentz-violating mass terms \cite{Rubakov:2004eb}.

The paper is organized as follows. In Section 2 we find the most general  action in a flat $D$-dimensional Minkowski spacetime action built by means of a symmetric rank-2 tensor $h_{\mu\nu}(x)$ invariant under the symmetry \eqref{1.1}. For this action, a gauge fixing term is written, which depends on two gauge parameters, and a mass term, which depends on two mass parameters as well. In Section 3 the setting to compute the propagators is prepared. The propagators of the theory are parametrized in terms of a basis of projector operators, and depend on a few coefficients, to be determined by solving four set of equations. In Section 4 the solutions of these equations are given. There are only four possible choices of the gauge fixing term, which lead to to well defined massive propagators. The condition of absence of unphysical poles is imposed, which, for each solution, translates into constraints on the two mass parameters. We find that the FP mass term is not \textcolor{black}{the unique choice}, and the massless limit of the various solutions is studied, with the outcome that {two solutions} have a smooth limit. Remarkably, we find that the FP case is, indeed, peculiar, since only in this case it is possible to define a nilpotent BRS-like operator which is a symmetry  and whose unique element of the cohomology is just the FP massive theory of gravity. Our conclusions are summarized in Section 5.

\section{Lorentz-invariant \textcolor{black}{mass term}}

The most general $D$-dimensional action for a rank-2 symmetric tensor $h_{\mu\nu}(x)=h_{\nu\mu}(x)$ invariant under \eqref{1.1} is 

\begin{equation}
S_{inv}=\int d^Dx
\left(
\frac{1}{2}h\partial^2h - h_{\mu\nu}\partial^\mu\partial^\nu h - \frac{1}{2}h^{\mu\nu}\partial^2h_{\mu\nu} + h^{\mu\nu}\partial_\nu\partial^\rho h_{\mu\rho}
\right),
\label{2.2}\end{equation}

where $h(x)=h^\lambda_\lambda(x)=\eta^{\mu\nu}h_{\mu\nu}(x)$, $\eta_{\mu\nu}=\mbox{Diag}(-1,1,\ldots,1)$ is the flat $D$-dimensional Minkowskian metric and the canonical mass dimension of $h_{\mu\nu}(x)$ is $[h_{\mu\nu}]=\frac{D-2}{2}$.

We can promote the gauge symmetry \eqref{1.1} to the BRS transformations 
\begin{eqnarray}
s h_{\mu\nu} &=& \partial_\mu\xi_\nu + \partial_\nu\xi_\mu \\
s \xi_\mu &=& 0 \\
s \bar\xi_\mu &=& b_\mu \\
s b_\mu &=&0,
\end{eqnarray}

where $\xi_\mu(x)$, $\bar\xi_\mu(x)$ and $b_\mu(x)$ are are ghost, antighost fields, and Lagrange multiplier, respectively. By means of the BRS transformations, which are trivially nilpotent $s^2=0$, an invariant gauge fixing term can be added to the action
\begin{eqnarray}
S_{gf} &=&s\int d^Dx\ \bar\xi^\mu\left(F_\mu(h) + \frac{\kappa}{2}b_\mu\right)\nonumber \\
&=& \int d^Dx \left[
b^\mu F_\mu(h)+\frac{\kappa}{2}b^\mu b_\mu +\partial^\nu\bar\xi^\mu \left( (1+2\kappa_1)\partial_\mu\xi_\nu+\partial_\nu\xi_\mu \right)
\right]\label{2.7}
\end{eqnarray}

{As in any gauge field theory, the role of the Lagrange multiplier (or Lautrup-Nakanishi field \cite{Lautrup:1967zz, Nakanishi:1966zz}) $b^{\mu}(x)$ is to enforce the gauge condition 
\begin{equation}
F_\mu(h)\equiv\partial^\nu h_{\mu\nu} +\kappa_1\partial_\mu h=0,\label{gaugecond}
\end{equation}
where $h\equiv h^\lambda_{\ lambda}$ is the trace of $h_{\mu\nu}$.

Indeed, after eliminating the lagrange multiplier $b^{\mu}(x)$ by means of its equation of motion, the gauge fixing term in the action reads
\begin{equation}
S_{gf}=\int d^Dx\left(-\frac{1}{2\kappa}F^{\mu}F_{\mu}\right) + \mbox{ghost sector}.\label{2.8}
\end{equation}
This is in complete analogy with ordinary gauge field theories built by means of the gauge field $A_{\mu}(x)$. like Maxwell or Yang-Mills theory in 4D, or topological Chern-Simons theory in 3D, where the gauge condition, instead of \eqref{gaugecond}, is given by the more familiar $F(A)=\partial_{\mu}A^{\mu}=0$. Pursuing that analogy, we might call the $\kappa=0$ and $\kappa=1$ cases Landau and Feynman gauge, respectively.
}

%

In \eqref{2.7},  $\kappa,\kappa_1$ are two gauge parameters. Notice that, in principle, we also allow the possibility for the gauge fixing of $h(x)$, $i.e.$ of the minkovskian trace of $h_{\mu\nu}(x)$. We shall see that this will turn out to be crucial for the FP case.

A mass term can be added to the action, whose most general, Lorentz invariant, form is 
\begin{equation}
S_m=\int d^Dx \frac{1}{2} \left (
m_1^2h^{\mu\nu}h_{\mu\nu} + m_2^2 h^2
\right).\label{2.9}
\end{equation}

The FP mass term corresponds to the particular case
\begin{equation}
\mbox{FP}: m_1^2+m_2^2=0
\label{2.10}\end{equation}

The mass term breaks the BRS symmetry of the total action $S=S_{inv}+S_{gf}+S_m$ as follows
\begin{equation}
sS=\int d^Dx\ \left(m_1^2\textcolor{black}{h^{\mu\nu}}(\partial_\mu\xi_\nu+\partial_\nu\xi_\mu)+2m_2^2\textcolor{black}{h}\partial^\mu\xi_\mu\right).
\label{2.11}\end{equation}

\section{Computing the propagators}

In order to compute the propagators of the theory in momentum space, it is convenient to introduce the rank-2 projectors

\begin{equation}
d_{\mu\nu} = \eta_{\mu\nu}-\frac{p_\mu p_\nu}{p^2}\ ;\
e_{\mu\nu} = \frac{p_\mu p_\nu}{p^2}, 
\label{3.1}\end{equation}

which are idempotent and orthogonal

\begin{equation}
d_{\mu\rho}d^\rho_\nu = d_{\mu\nu}\ ;\
e_{\mu\rho}e^\rho_\nu = e_{\mu\nu}\ ;\
d_{\mu\rho}e^\rho_\nu = e_{\mu\rho}d^\rho_\nu =0.
\label{3.2}\end{equation}

In terms of $d_{\mu\nu}$ and $e_{\mu\nu}$, it is possible to define the following four rank-4 projectors

\begin{eqnarray}
A_{\mu\nu,\rho\sigma} &=& \frac{d_{\mu\nu}d_{\rho\sigma}}{D-1} \label{3.3}\\
B_{\mu\nu,\rho\sigma} &=& e_{\mu\nu}e_{\rho\sigma}\label{3.4}\\
C_{\mu\nu,\rho\sigma} &=& \frac{1}{2} (d_{\mu\rho}d_{\nu\sigma}+d_{\mu\sigma}d_{\nu\rho}-\frac{2}{D-1}d_{\mu\nu}d_{\rho\sigma})\label{3.5}\\
D_{\mu\nu,\rho\sigma} &=& \frac{1}{2} (d_{\mu\rho}e_{\nu\sigma}+d_{\mu\sigma}e_{\nu\rho}+d_{\nu\rho}e_{\mu\sigma}+d_{\nu\sigma}e_{\mu\rho})\label{3.6},
\end{eqnarray}

which display the following symmetry properties
\begin{equation}
X_{\mu\nu,\rho\sigma}=X_{\nu\mu,\rho\sigma}=X_{\mu\nu,\sigma\rho}=X_{\rho\sigma,\mu\nu}\ ;\ X=\{A,B,C,D\},
\label{3.7}\end{equation}
and are idempotent and orthogonal as well

\begin{eqnarray}
A\cdot B=A\cdot C=A\cdot D=B\cdot C=B\cdot D=C\cdot D &=&0\\
B\cdot A=C\cdot A=D\cdot A=C\cdot B=D\cdot B=D\cdot C &=& 0\\
A\cdot A=A\ ;\ B\cdot B=B\ ;\ C\cdot C=C\ ;\ D\cdot D&=&D,
\end{eqnarray}

where $A\cdot B \equiv A^{\mu\nu,\alpha\beta}B_{\alpha\beta,\rho\sigma}$. The projectors $A,B,C$ and $D$ decompose the rank-4 identity $\mathbb{I}\equiv\frac{1}{2}(\eta_{\mu\rho}\eta_{\nu\sigma}+\eta_{\mu\sigma}\eta_{\nu\rho})$

\begin{equation}
A+B+C+D=\mathbb{I}.
\label{3.11}\end{equation}

In momentum space, the massive, gauge fixed action $S$ reads

\begin{equation}
S=\int d^Dp \left(
\tilde{h}_{\alpha\beta}\tilde{h}_{\gamma\delta}\Omega^{\alpha\beta,\gamma\delta}(p) +
\tilde{h}_{\alpha\beta}\tilde{b}_\nu\Lambda^{\alpha\beta,\nu}(p) +
\tilde{b}_\mu\tilde{b}_\nu H^{\mu\nu}(p) 
\right),
\label{3.12}\end{equation}

where the functions $\Omega(p), \Lambda(p)$ and $H(p)$ can be written in terms of the projection operators \eqref{3.3}, \eqref{3.4}, \eqref{3.5} and \eqref{3.6} as

\begin{equation}
\Omega^{\alpha\beta,\gamma\delta}(p) = tA^{\alpha\beta,\gamma\delta}+uB^{\alpha\beta,\gamma\delta}+
vC^{\alpha\beta,\gamma\delta}+zD^{\alpha\beta,\gamma\delta},
\label{3.13}\end{equation}

with

\begin{eqnarray}
t &=& \frac{2-D}{2}p^2+\frac{1}{2}(m_1^2+Dm_2^2) \label{3.14}\\
u &=& \frac{1}{2}(m_1^2+Dm_2^2)  \label{3.15}\\
v &=& \frac{1}{4}(p^2+m_1^2) \label{3.16}\\
z &=& \frac{1}{2}m_1^2 \label{3.17}
\end{eqnarray}
and

\begin{eqnarray}
\Lambda^{\alpha\beta,\nu} &=& 
-i\left[
\frac{1}{2}(d^{\alpha\nu} p^\beta + d^{\beta\nu} p^\alpha)+\kappa_1 d^{\alpha\beta}p^\nu+(1+\kappa_1)e^{\alpha\beta}p^\nu\right] \label{3.18}\\
H^{\mu\nu} &=&
\frac{\kappa}{2}(d^{\mu\nu}+e^{\mu\nu}).\label{3.19}
\end{eqnarray}

Let us parametrize the propagators 

\begin{eqnarray}
\langle \tilde{h}_{\mu\nu}\tilde{h}_{\rho\sigma} \rangle (p) &=& 
G_{\mu\nu,\rho\sigma} (p)\label{3.20}\\
\langle \tilde{h}_{\mu\nu}\tilde{b}_\rho \rangle(p) &=& G_{\mu\nu,\rho} (p)\label{3.21}\\
\langle \tilde{b}_\mu\tilde{b}_\nu \rangle(p) &=& G_{\mu\nu} (p)\label{3.22}
\end{eqnarray}

as follows

\begin{eqnarray}
G_{\mu\nu,\rho\sigma} (p) &=& \hat{t}A_{\mu\nu,\rho\sigma}+\hat{u}B_{\mu\nu,\rho\sigma}+
\hat{v}C_{\mu\nu,\rho\sigma}+\hat{z}D_{\mu\nu,\rho\sigma} \label{3.23}\\
G_{\mu\nu,\rho} (p) &=& if(p_\mu d_{\nu\rho} + p_\nu d_{\mu\rho} )
+igp_\rho d_{\mu\nu}
+i\textcolor{black}{l}p_\rho e_{\mu\nu}\label{3.24}\\
G_{\mu\nu} (p) &=&rd_{\mu\nu}+se_{\mu\nu},\label{3.25}
\end{eqnarray}

where the set of constants $\{\ \hat{t}\ ; \hat{u}\ ; \hat{v}\ ; \hat{z}\ ; f\ ; g\ ; \textcolor{black}{l}\ ; r\ ; s\ \}$ must be determined by the request that the matrix of propagators is the inverse of the quadratic action \eqref{3.12}, which translates into the equations

\begin{eqnarray}
\Omega^{\mu\nu,\alpha\beta}G_{\alpha\beta,\rho\sigma} +\Lambda^{\mu\nu,\alpha}(G^*)_{\alpha,\rho\sigma} &=& \mathbb{I}^{\mu\nu}_{\rho\sigma} \label{3.26}\\
\Omega^{\mu\nu,\alpha\beta}G_{\alpha\beta,\rho} +\Lambda^{\mu\nu,\alpha}G_{\alpha\rho} &=& 0 \label{3.27}\\
(\Lambda^*)^{\mu,\gamma\delta}G_{\gamma\delta,\rho\sigma} +H^{\mu\alpha}(G^*)_{\alpha,\rho\sigma} &=&0 \label{3.28}\\
(\Lambda^*)^{\mu,\gamma\delta}G_{\gamma\delta,\sigma} +H^{\mu\alpha}G_{\alpha\sigma} &=&\mathbb{I}^{\mu}_{\sigma} \label{3.29}
\end{eqnarray}

The above four equations give rise to

\begin{equation}\eqref{3.26}\Rightarrow
\left\{
\begin{array}{rcc}
t\hat{t}-\frac{\kappa_1}{2}(D-1) gp^2&=& 1 \\
u\hat{u}-\frac{1+\kappa_1}{2}\textcolor{black}{l}p^2 &=& 1 \\
v\hat{v} &=& 1 \\
z\hat{z}-fp^2 &=& 1 \\
\kappa_1\textcolor{black}{l}+(1+\kappa_1)g &=&0
\end{array}
\right.
\label{3.30}\end{equation}

\begin{equation}\eqref{3.27}\Rightarrow
\left\{
\begin{array}{rcc}
fz-\frac{1}{4}r &=& 0 \\
\textcolor{black}{l}u-\frac{1+\kappa_1}{2}s &=& 0\\
gt-\frac{\kappa_1}{2}s &=& 0
\end{array}
\right.
\label{3.31}\end{equation}

\begin{equation}\eqref{3.28}\Rightarrow
\left\{
\begin{array}{rcc}
\frac{1}{2}\hat{z}-\kappa f &=& 0 \\
\kappa_1\hat{t}-\kappa g &=& 0 \\
(1+\kappa_1)\hat{u}-\kappa \textcolor{black}{l} &=& 0
\end{array}
\right.
\label{3.32}\end{equation}

\begin{equation}\eqref{3.29}\Rightarrow
\left\{
\begin{array}{rcc}
\kappa r -fp^2&=& 2 \\
\kappa s -\kappa_1(D-1)gp^2-(1+\kappa_1)\textcolor{black}{l}p^2 &=& 2,
\end{array}
\right.
\label{3.33}\end{equation}

which must be solved in order to find all possible propagators of the theory. 

\section{Solutions \textcolor{black}{and massless limit}}

In this section we list all possible solutions of the four sets of equations \eqref{3.30}. \eqref{3.31}, \eqref{3.32} and \eqref{3.33}, for which propagators do exist. For each solution, we shall find the range for the masses $m_1^2$ and $m_2^2$ which guarantees the absence of unphysical poles.

We classify the solutions in terms of the gauge parameters $\kappa$ and $\kappa_1$, and it turns out that for the latter  only two possible values are allowed: 
\begin{equation}
\kappa_1=0;-1.
\label{4.1}\end{equation}

\textcolor{black}{As it is well known, linearized gravity is a gauge theory (see for instance Sections 7.2 and 7.3 of \cite{Carroll:2004st}) for which several possible choices of gauge fixing exist. Amongst these, a covariant choice is the Lorenz/harmonic one $\partial_\mu h^\mu_{\ \nu}-\frac{1}{2}h=0$ (eq (7.46) of \cite{Carroll:2004st}), which involves the trace $h=h^\lambda_{\ \lambda}$ of the metric perturbation $h^\mu_{\ \nu}$ . It is quite remarkable that imposing the absence of propagating tachyons, we find the constraint (4.1) on the gauge fixing condition (2.7), which implies only two possibilities for the gauge fixing term: the Landau gauge and the  ``Lorenz/harmonic like'' gauge, which is of the same type of (7.46) of  \cite{Carroll:2004st}. Hence, one of the first principles of quantum field theory (absence of tachyonic propagators) strictly selects the possible gauge conditions, which therefore apparently are not left to free choice.
}

For each of the two possible values of $\kappa_1$, we separately analyze the cases $\kappa=0$, which corresponds to the Landau gauge, and $\kappa\neq 0$. Hence we have four possibilities.

\begin{enumerate}
\item 
\begin{equation}\kappa\neq 0\ ;\ \kappa_1=0:\label{4.2}\end{equation}

we get
\begin{equation}
f=\frac{2}{2\kappa m_1^2-p^2}  \ ;\  g=0 \ ;\      \textcolor{black}{l}=\frac{2}{\kappa (m_1^2+Dm_2^2)-p^2}
\label{4.3}\end{equation}
\begin{equation}
r=\frac{4m_1^2}{2\kappa m_1^2-p^2}      \ ;\  s=\frac{2(m_1^2+Dm_2^2)}{\kappa (m_1^2+Dm_2^2)-p^2}
\label{4.4}\end{equation}
\begin{eqnarray}
\hat{t}=\frac{2}{(2-D)p^2+m_1^2+Dm_2^2} &\ ;\ &
\hat{u}=\frac{2\kappa}{\kappa (m_1^2+Dm_2^2)-p^2} \label{4.5}\\
\hat{v}=\frac{4}{p^2+m_1^2} &\ ;\ &
\hat{z}=\frac{4\kappa}{2\kappa m_1^2-p^2}.\label{4.6}
\end{eqnarray}

Imposing the absence of \textcolor{black}{unphysical poles} leads to
\begin{equation}
m_1^2\geq 0\ ;\
m_1^2+Dm_2^2\leq 0\ ;\
\kappa m_1^2\leq 0\ ;\
\kappa (m_1^2+Dm_2^2)\leq 0,
\label{4.7}\end{equation}

hence
{
\begin{description}
\item[solution 1] 
\begin{equation}
m_1^2> 0 \ ;\ m_1^2+Dm_2^2= 0\ ;\ \kappa<0\ ;\ \kappa_1=0.
\label{4.8}\end{equation}
\end{description}
}
{
A simple example for this solution (putting $\kappa=-\frac{1}{2}$ and $D=4$) reads 
\begin{equation}
f=-\frac{2}{p^2+m_1^2}  \ ;\  g=0 \ ;\      \textcolor{black}{l}=-\frac{2}{p^2}
\label{4.9}\end{equation}
\begin{equation}
r=-\frac{4m_1^2}{p^2+m_1^2}      \ ;\  s=0
\label{4.10}\end{equation}
\begin{equation}
\hat{t}=-\frac{1}{p^2} \ ;\ 
\hat{u}=\frac{1}{p^2} \label{4.11}\ ;\
\hat{v}=\frac{4}{p^2+m_1^2} \ ;\ 
\hat{z}=\frac{2}{p^2+m_1^2}.
\end{equation}
}
\item \begin{equation}\kappa\neq 0\ ;\ \kappa_1=-1:\label{4.13}\end{equation}

we get
\begin{equation}
f=\frac{2}{2\kappa m_1^2-p^2}  
;
g=-\frac{2}{[2\kappa+1-D(\kappa+1)]p^2+\kappa(m_1^2+Dm_2^2)} 
;  
\textcolor{black}{l}=0
\label{4.14}\end{equation}
\begin{equation}
r=\frac{4m_1^2}{2\kappa m_1^2-p^2} 
;
s=\frac{2[(2-D)p^2+m_1^2+Dm_2^2]}{[2\kappa+1-D(\kappa+1)]p^2+\kappa(m_1^2+Dm_2^2)} 
\label{4.15}\end{equation}
\begin{eqnarray}
\hat{t}&=&\frac{2\kappa}{[2\kappa+1-D(\kappa+1)]p^2+\kappa(m_1^2+Dm_2^2)} \label{4.16}\\
\hat{u}&=&\frac{2}{m_1^2+Dm_2^2} ;
\hat{v}=\frac{4}{p^2+m_1^2} 
;
\hat{z}=\frac{4\kappa}{2\kappa m_1^2-p^2}.   \label{4.17}
\end{eqnarray}

As in the previous case, imposing the absence of unphysical poles in the coefficients which appear in the propagators \eqref{3.23}, \eqref{3.24} and \eqref{3.25}, we find five non trivial possibilities, hence five distinct solutions: 
\begin{description}
\item[solution 2] 
\begin{equation}
m_1^2>0\ ;\ 
m_1^2+Dm_2^2<0\ ;\ 
\kappa\leq\frac{D-1}{2-D}\ ;\ \kappa_1=-1
\label{4.18}\end{equation}
\item[solution 3] 
\begin{equation}
m_1^2>0\ ;\ 
m_1^2+Dm_2^2>0\ ;\ 
\frac{D-1}{2-D}\leq\kappa<0\ ;\ \kappa_1=-1
\label{4.19}\end{equation}
\item[solution 4] 
\begin{equation}
m_1^2=0\ ;\
m_2^2<0\ ;\ \kappa\leq\frac{D-1}{2-D}\ ;\ \kappa_1=-1
\label{4.20}\end{equation}
\item[solution 5] 
\begin{equation}
m_1^2=0\ ;\
m_2^2>0\ ;\ 
\frac{D-1}{2-D}\leq\kappa<0\ ;\ \kappa_1=-1
\label{4.21}\end{equation}
\item[solution 6] 
\begin{equation}
m_1^2=0\ ;\
m_2^2<0\ ;\ \kappa>0\ ;\ \kappa_1=-1
\label{4.22}\end{equation}
\end{description}

{
Simple examples can be obtained also for this case \label{4.13}. For instance, putting $\kappa=-\frac{1}{2}$, $D=4$ and $m_2^2=\frac{3}{4}m_1^2$, choice which belongs to solution 3 \eqref{4.19}, we get
\begin{equation}
f=-\frac{2}{p^2+m_1^2}  \ ;\
g=\frac{1}{p^2+m_1^2} \ ;\
\textcolor{black}{l}=0
\end{equation}
\begin{equation}
r=-\frac{4m_1^2}{p^2+m_1^2} \ ;\
s=\frac{2(p^2-m_1^2)}{p^2+m_1^2} 
\end{equation}
\begin{equation}
\hat{t}=\frac{1}{2(p^2+m_1^2)} \ ;\
\hat{u}=\frac{1}{2m_1^2} \ ;\
\hat{v}=\frac{4}{p^2+m_1^2} \ ;\
\hat{z}=\frac{2}{p^2+m_1^2}.   
\end{equation}
Similarly, if in solution 5 \eqref{4.21}, where $m_1^2=0$, we put $\kappa=-\frac{1}{2}$, and $D=4$, 
we get
\begin{equation}
f=-\frac{2}{p^2}  \ ;\
g=-\frac{1}{p^2+m_2^2} \ ;\
\textcolor{black}{l}=0
\end{equation}
\begin{equation}
r=0\ ;\
s=\frac{2(p^2-2m_2^2)}{p^2+m_2^2} 
\end{equation}
\begin{equation}
\hat{t}=\frac{1}{2(p^2+m_2^2)} \ ;\
\hat{u}=\frac{1}{2m_2} \ ;\
\hat{v}=\frac{4}{p^2} \ ;\
\hat{z}=\frac{2}{p^2}.   
\end{equation}
Alternatively, choosing the extremal value for $\kappa=\frac{D-1}{2-D}$, which is allowed in solutions 2,3,4 and 5, it is possible to switch off the degree of freedom related to the coefficient $\hat{t}$ in the $\textcolor{black}{l}$-sector of the propagators, together with the one related to $\hat{u}$, which does not propagate ever.}

\item \begin{equation}\kappa=0\ ;\ \kappa_1=0:\label{4.23}\end{equation}

we get
\begin{equation}
f=-\frac{2}{p^2}  \ ;\  g=0 \ ;\  \textcolor{black}{l}=-\frac{2}{p^2}
\label{4.24}\end{equation}
\begin{equation}
r=-\frac{4m_1^2}{p^2} \ ;\  s=-\frac{2}{p^2}(m_1^2+Dm_2^2)
\label{4.25}\end{equation}
\begin{equation}
\hat{t}=\frac{2}{(2-D)p^2+m_1^2+Dm_2^2}\ ;\ 
\hat{u}=0 \ ;\  
\hat{v}=\frac{4}{p^2+m_1^2} \ ;\   
\hat{z}=0.
\label{4.26}\end{equation}

Imposing the absence of \textcolor{black}{tachyonic poles}, we find
\begin{description}
\item[solution 7] 
\begin{equation}
m_1^2 \geq 0 \ ;\ m_1^2+Dm_2^2 \leq 0\ ;\ \kappa=0\ ;\ \kappa_1=0
\label{4.27}\end{equation}
\end{description}

For instance, at the point $m_1^2+Dm_2^2=0$ we have

\begin{equation}
f=-\frac{2}{p^2}  \ ;\  g=0 \ ;\  \textcolor{black}{l}=-\frac{2}{p^2}
\end{equation}
\begin{equation}
r=-\frac{4m_1^2}{p^2} \ ;\  s=0
\end{equation}
\begin{equation}
\hat{t}=\frac{2}{(2-D)p^2}\ ;\ 
\hat{u}=0 \ ;\  
\hat{v}=\frac{4}{p^2+m_1^2} \ ;\   
\hat{z}=0.
\end{equation}

\item \begin{equation}\kappa=0\ ;\ \kappa_1=-1:\label{4.28}\end{equation}
we get
\begin{equation}
f=-\frac{2}{p^2}  
\ ;\ 
g=\frac{2}{(D-1)p^2} 
\ ;\     
\textcolor{black}{l}=0
\label{4.29}\end{equation}
\begin{equation}
r=-\frac{4m_1^2}{p^2} 
\ ;\
s=\frac{2[(2-D)p^2+m_1^2+Dm_2^2]}{(1-D)p^2} 
\label{4.30}\end{equation}
\begin{equation}
\hat{t}=0
\ ;\
\hat{u}=\frac{2}{m_1^2+Dm_2^2} 
\ ;\
\hat{v}=\frac{4}{p^2+m_1^2} 
\ ;\
\hat{z}=0.
\label{4.31}\end{equation}

Imposing the absence of unphysical poles, we find

{
\begin{description}
\item[solution 8] 
\begin{equation}
m_1^2\geq 0 \ ;\ m_1^2+Dm_2^2\neq 0\ ;\ \ ;\ \kappa=0\ ;\ \kappa_1=-1
\label{4.32}\end{equation}
\end{description}
}
\end{enumerate}


{We found eight solutions, each describing a different theory of massive gravity, free of tachyons (unphysical poles in the propagators). Amongst them, the solution 1 \eqref{4.8} and solution  7 \eqref{4.27}, $i.e$ the cases where the trace $h(x)$ is not gauge-fixed ($\kappa_1=0$), are peculiar, because they display a good massless limit $m_1^2,m_2^2\rightarrow 0$. For this reason, from the pure field theoretical point of view, these are acceptable under any respect. We remark that while the FP case belongs to the range \eqref{4.27} of solution 7, this is not the case for solution 1 \eqref{4.8}, which excludes the FP point.}

%
%
%
%

\subsection{Cohomological identification of the Fierz-Pauli case}

{Consider} the five solutions \eqref{4.18}, \eqref{4.19}, \eqref{4.20}, \eqref{4.21} and \eqref{4.22} which correspond to $\kappa\neq0$ and $\kappa_1=-1$. All of them \textcolor{black}{meet our requirements}    
and have a divergent massless limit. Amongst them, only the solution 2 \eqref{4.18} contains the FP case $m_1^2+m_2^2=0$, while the others represent massive theories of gravity which do not include the FP case. Notice that solutions 4 \eqref{4.20} , 5 \eqref{4.21} and 6 \eqref{4.22}, which do not contain the FP point, are characterized by one of the masses, namely $m_1^2$, set to zero. This situation closely resembles the solutions found in \cite{Rubakov:2004eb}.

On the other hand, solution 8 \eqref{4.32} corresponds to the Landau gauge with $\kappa=0$ and $\kappa_1=-1$, and has a divergent massless limit. The range of allowed masses contains the FP case, but, as we shall see shortly,  in this case (and only in this case) the latter can be isolated univocally. 

Up to now, the request that a massive theory of gravity is free \textcolor{black}{of unphysical poles} is not enough to single out the FP massive gravity. On the contrary, we found \textcolor{black}{admissible}   
solutions which exclude the FP case (solutions 1 \eqref{4.8}, 3 \eqref{4.19}, 4 \eqref{4.20}, 5 \eqref{4.21} and 6 \eqref{4.22}). {The solutions which distinguish from the others, so far, are solutions 1 \eqref{4.8} and 7 \eqref{4.27}, since these are the only ones to have a good massless limit, and the FP point is outside the range of solution 1, while it represents only one of the infinite possible choices of the two masses involved in solution 7.}

{Let us take a closer look to solution 8 \eqref{4.32}. In this case the FP mass term can be isolated by means of a nilpotent BRS symmetry, which univocally characterizes the FP theory of massive gravity as the only element belonging to the cohomology of a nilpotent BRS operator. This can be done as follows.}

The gauge fixing and mass terms corresponding to solution 8 \eqref{4.32} in the FP point $m_2^2=-m_1^2=-m^2$  are

\begin{eqnarray}
S_{gf} &=& \int d^Dx\ b^\mu(\partial^\nu h_{\mu\nu}-\partial_\mu h) \label{4.37}\\
S^{(FP)}_{m} &=&\frac{m^2}{2}\int d^Dx \left(h^{\mu\nu}h_{\mu\nu}-h^2\right).\label{4.38}
\end{eqnarray}

The mass term $S^{(FP)}_{m}$ breaks the BRS invariance of the action $S_{FP}\equiv S_{inv}+S_{gf}+S^{(FP)}_{m}$:

\begin{eqnarray}
sS_{FP}=sS^{(FP)}_m &=& m^2\int d^Dx\ h^{\mu\nu}(\partial_\mu \xi_\nu +\partial_\nu \xi_\mu - 2h\partial^\mu\xi_\mu)\nonumber\\
&=& -2m^2\int d^Dx\ \xi_\mu(\partial_\nu h^{\mu\nu}-\partial^\mu h),\label{4.39}
\end{eqnarray}

but, thanks to the particular gauge fixing term \eqref{4.37} peculiar to the solution 8, the breaking induced by the FP mass term can be reabsorbed as follows
\begin{equation}
sS_{FP}=-2m^2\int d^Dx\ \xi_\mu\frac{\delta S_{FP}}{\delta b_\mu}.
\label{4.40}\end{equation}

This suggests to define the modified symmetry operator $\hat{s}$

\begin{eqnarray}
\hat{s} h_{\mu\nu} &=& \partial_\mu\xi_\nu + \partial_\nu\xi_\mu \\
\hat{s} \xi_\mu &=& 0 \\
\hat{s} \bar\xi_\mu &=& b_\mu \\
\hat{s} b_\mu &=&2m^2\xi_\mu,
\end{eqnarray}

in terms of which the gauge fixing term can be written

\begin{equation}
S_{gf}=\hat{s}\int d^Dx\ \bar\xi^\mu (\partial^\nu h_{\mu\nu}-\partial_\mu h).
\label{4.45}\end{equation}

This new operator describes a symmetry of the gauge fixed, $massive$ whole action $S_{FP}$

\begin{equation}
\hat{s}S_{FP}=0,
\label{4.46}\end{equation}

but $\hat{s}$ is not a BRS operator, since it is not nilpotent (even if this happens only on the antighost sector)~: $\hat{s}^2\bar\xi_\mu=2m^2\xi_\mu$. 

Once we define 

\begin{eqnarray}
\delta h_{\mu\nu} &=& \delta\xi_\mu=\delta b_\mu=0 \\
\delta\bar\xi_\mu &=& 2m^2\xi_\mu,
\end{eqnarray}

the symmetry operators $\hat{s}$ and $\delta$ form the algebra

\begin{equation}
\hat{s}^2=\delta\ ;\ [\hat{s},\delta]=0,
\label{4.49}\end{equation}

which closely resembles the one characterizing the much more complicated supersymmetric gauge field theories \cite{Maggiore:1994xw,Maggiore:1994dw}. Following the strategy adopted there, we introduce a constant anticommuting ghost $\epsilon$, with Faddeev-Popov charge $-1$, and we define the operator

\begin{equation}
Q\equiv\hat{s}+\epsilon\delta-\frac{\partial}{\partial\epsilon},
\label{4.50}\end{equation}

which describes a symmetry of the whole massive theory

\begin{equation}
{\cal Q}S_{FP}=0,
\label{4.51}\end{equation}

and, thanks to the algebra \eqref{4.49}, is nilpotent, which makes ${\cal Q}$ a true BRS operator

\begin{equation}
{\cal Q}^2=0.
\label{4.52}\end{equation}

At this point it is easy to verify, using standard techniques \cite{Piguet:1995er}, that the FP action for massive gravity $S_{FP}$ is the only element belonging to the cohomology of the BRS operator ${\cal Q}$, which renders the FP case definitely peculiar from the field theoretical point of view.

\section{Conclusions}

\textcolor{black}{In this paper we considered the problem of adding the most general mass term, which depends of two mass parameters $m_1^2$ and $m_2^2$, to the gauge field theory of a rank-2 symmetric tensor $h_{\mu\nu}(x)$, whose transformation law is given by (1.1).  
This issue is complementary to the study of the gauge field theory of antisymmetric tensor fields \cite{Deser:1970dqa, Kalb:1974yc,Freedman:1980us}. In that cases, one lands on the study of a kind of topological field theories, called BF theories \cite{Birmingham:1991ty}, which have the peculiar property of being finite, $i.e.$ without divergent quantum corrections \cite{Maggiore:1991aa,Maggiore:1992ug}. 
It is therefore natural to extend the same kind of study to the theory of D symmetric tensor field, which, together with the transformation law (1.1), gives rise to the action (2.1).
We stress that our point of view is, in a sense, from below: instead of considering the action (2.1) as the the starting point as the linearised version of the EH action, we see it as the most general action invariant under the transformation (1.1) involving a symmetric tensor field. 
The bottom line is that the two approaches coincide, but our perspective is more from the quantum field theory side rather than from pure GR one. 
This might help to understand our road map: as the preliminary step, we faced the problem of fixing the gauge. This task is a highly non trivial one for the theory of antisymmetric tensor fields. The gauge transformations in that case are reducible, and ghost for ghost fields are needed, for which the Batalin - Vilkovisky machinery must be used \cite{Batalin:1981jr,Batalin:1984jr}. 
The case of the D symmetric tensor theory, is far less complicated, although some care should be taken also in this case due to the possibility of fixing the gauge also for the trace $h(x)$ of $h_{\mu\nu}(x)$. Therefore  the symmetry which we gauge fixed is (1.1), as it should. To do this, we adopted the standard BRS procedure, by first promoting the gauge parameter appearing in (1.1) to ghost field, and then adding to the action (1.1) the most general BRS cocycle (2.6), which involves the usual gauge field team: gauge field, ghost and antighost anticommuting fields, and finally the Lagrange multiplier (or Lautrup-Nakanishi field) $b_{\mu}(x)$ 
In the usual BRS approach, it is the Lagrange multiplier which enforces the gauge condition (2.7). As the ghost/antighost fields, it appears in the gauge fixing term, which, being a BRS cocycle, is not physical. Its presence renders the propagators matrix non diagonal, and most part of this paper has been in fact devoted to computing the whole matrix of propagators, formed by (3.20), (3.21) and (3.22).
At this point, the gauged fixed action is classically well defined, and the (exact) symmetry involved is the BRS one. After this,  we added the most general, covariant, mass term (2.8), which, of course breaks the BRS symmetry of the gauge fixed action. We stress that we add the mass term $after$ having gauge fixed the action, as it must be done in an ordinary gauge field theory, not before. The breakings due to mass terms are of great importance in gauge field theory, and trying to master the breakings is of primary importance. We tuned the two free coefficients appearing in the mass term we introduced having in mind the classification of the least harming mass terms in supersymmetric gauge field theories, where the introduction of mass terms for the supersymmetric partners of ordinary particles is mandatory. In that context, the pioneering result is that due to Girardello and Grusaru \cite{Girardello:1981wz}, who were able to identify soft supersymmetric breaking mass terms. Their result stands to N=1 SYM theory as the FP mass term stands to the linearised EH action of GR.
We followed the same procedure, by asking as necessary condition for an acceptable mass broken field theory, the absence of unphysical poles of tachyonic type in the (whole matrix of) propagators. 
%
This requirement led to constraints on the four parameters (2 gauge, 2 masses) on which the gauge fixed theory with breaking mass terms depends.} 
\textcolor{black}{We emphasize that the absence of unphysical poles is a necessary but not sufficient requirement to identify a physical model. The analysis of this point is scheduled for a forthcoming paper.
}

\textcolor{black}{In our paper we discuss the eight possibilities which turn out. Amongst these, some have good massless limit, some have not, but all of them are ``soft broken massive theories'', in the sense of Girardello and Grisaru, and hence potentially acceptable.
%
We found the existence of several (eight) soft breaking mass terms for the gauge fixed theory of a D symmetric tensor field. }

The main results presented in this paper are the following

\begin{enumerate}
\item
The only possible gauge fixing terms leading to well defined massive theories, correspond to 
{four} choices of the gauge parameters: 
	\begin{enumerate}
	\item $\kappa\neq 0, \ \kappa_1=0$
	\item $\kappa\neq 0, \ \kappa_1=-1$
	\item $\kappa = 0, \ \kappa_1=0$
	\item $\kappa = 0, \ \kappa_1=-1$
	\end{enumerate}
\item
Requesting the absence of unphysical poles in the propagators, we find 
{eight} disjoint sets for the masses $m_1^2$ and $m_2^2$
	\begin{enumerate}
	\item solution 1 \eqref{4.8} for $\kappa\neq 0, \ \kappa_1=0$
	\item solution 2 \eqref{4.18}, solution 3 \eqref{4.19}, solution 4 \eqref{4.20}, solution 5 \eqref{4.21} 
	and solution 6 \eqref{4.22} for $\kappa\neq 0,\ \kappa_1=-1;$
	\item solution 7 \eqref{4.27} for $\kappa= 0,\ \kappa_1=0$
	\item solution 8 \eqref{4.32} for $\kappa= 0,\ \kappa_1=-1.$
	\end{enumerate}
\item
The constraint of absence of tachyons alone is not sufficient to isolate the FP case $m_1^2+m_2^2=0$. There are solutions which exclude the FP point, namely solution 1 \eqref{4.8}, solution 3 \eqref{4.19}, solution 4 \eqref{4.20}, solution 5 \eqref{4.21} and solution 6 \eqref{4.22}.
\item
Two solutions display a good massless limit $m_1^2,m_2^2\rightarrow 0$: solution 1 \eqref{4.8} and solution 7 \eqref{4.27}, for which the trace $h$ is not gauge-fixed  ($\kappa_1=0$). Amongst these, only the solution 7 includes the FP point.
\item
Amongst the solutions we found, our approach gives a strong argument in favour of the standard FP term: the solution 8 \eqref{4.32} is the one for which the FP case can be singled out, by means of the BRS operator ${\cal Q}$ \eqref{4.50}, which describes a symmetry of the whole massive, gauged fixed action $S_{FP}$. The corresponding gauge fixing term involves also the trace $h(x)$ of $h_{\mu\nu}(x)$.
\textcolor{black}{We believe that this simple field theoretical argument in favour of the  FP massive gravity ($i.e.$ that the D linearised EH action with FP mass term is the only term belonging to the cohomology of a BRS operator) might be appreciated by the GR community. The remarkable fact, which, from the theoretical fact renders the FP case unique, is that the FP massive gravity theory is the only element of the cohomology of the BRS operator ${\cal Q}$.}
\end{enumerate}
\textcolor{black}{Finally, we stress once again the we added the mass terms $after$ having gauge fixed the action. Therefore, in order to tell if any of the solutions presented in this paper is somehow related to the FP theory of massive gravity, a careful counting of the degrees of freedom should be performed, together with the presence, or the absence, of the Boulware-Deser ghost \cite{Boulware:1973my}, whose absence uniquely identifies the FP mass terms in the linearized EH case. We believe that the results presented in our paper motivate further studies in that direction.}

{\bf Acknowledgements}

We would like to thank Andrea Amoretti, Matteo Baggioli and Nico Magnoli for friendly and helpful discussions.  N.M. thanks the support of INFN Scientific Initiative SFT: ``Statistical Field Theory, Low-Dimensional Systems, Integrable Models and Applications''


\end{document}